\documentclass[pra,twocolumn,showpacs,amsmath,amssymb,superscriptaddress]{revtex4}

\usepackage{graphicx}
\usepackage{dcolumn}
\usepackage{bm,color}

\begin{document}

\title{Scattering of polarized laser light by an atomic gas in free space: a QSDE approach}

\author{Luc Bouten} 
\author{John Stockton}
\author{Gopal Sarma}
\author{Hideo Mabuchi}
\affiliation{Physical Measurement and Control 266-33, California
Institute of Technology, Pasadena, CA 91125}


\begin{abstract}
We propose a model, based on a quantum stochastic 
differential equation (QSDE), to describe 
the scattering of polarized laser light by an atomic gas. 
The gauge terms in the QSDE account for the direct 
scattering of the laser light into different field channels. 
Once the model has been set, we can rigorously derive  
quantum filtering equations for balanced polarimetry and 
homodyne detection experiments, study the statistics of 
output processes and investigate a  
strong driving, weak coupling limit.  
\end{abstract}

\maketitle

\section{Introduction}

Many recent experimental 
\cite{SCJ03, JSCFP04, SCSDJ04, Sto06}
and theoretical \cite{THTTIY99, TMW02a, TMW02b, SiD03, GeB06, ShM06}
works have been based on a simple experimental 
scenario, in which a polarized atomic gas is 
continuously probed with a polarized off-resonant 
optical beam (Fig.\ \ref{Fig one}). By measuring 
the Faraday rotation of the optical polarization 
resulting from the interaction, one can in 
principle prepare conditionally spin-squeezed 
states or perform quantum metrology tasks, 
e.g.\ estimating a magnetic field that rotates 
the spins.

Central to the description of these
experiments is the \emph{quantum filtering equation}, 
which propagates the expectation value of 
the atomic gas observables conditioned on prior 
measurement results. The conditional 
expectation is the mean least squares estimate 
of an atomic gas observable given the observations 
thus far. The conditional expectations of `all' 
atomic observables can be summarized in an 
information state $\pi_t$. The filtering 
equation propagates this information state in 
real time. 

In quantum optics the filtering equation is 
often referred to as the \emph{stochastic master 
equation} \cite{Car93}. For the polarimetry 
example considered here, previous modelling 
efforts have either produced an unconditional 
description \cite{SiD03} or arrived at a 
conditional description by heuristically 
`adding the usual measurement terms' \cite{Sto06} 
in analogy with a physicallly different homodyne 
measurement scheme with only a single polarization 
mode \cite{TMW02a,TMW02b}. In this article 
we treat the conditional evolution of the 
state (due to detection of Faraday rotation 
with a polarimeter) in a rigorous manner, 
allowing the atomic system to mediate exchange 
between two orthogonal optical polarization 
modes. In particular, we derive the 
quantum filtering equation from an underlying 
quantum stochastic model, i.e.\ the 
quantum stochastic differential equation (QSDE) 
governing the interaction of the atomic gas 
with the laser light.  

Formal quantum filtering theory was pioneered 
by V.P.\ Belavkin in \cite{Bel88,Bel92b}
using martingale techniques (see also \cite{BGM04}). 
We here employ the reference probability method, 
based on the quantum Bayes formula \cite{BVH05a,BVH05b}, 
to obtain the quantum filter from the QSDE 
(see also \cite{Hol90}). 

The QSDE model we use here is based on a simple 
Faraday Hamiltonian, $H = \kappa F_zS_z$, where 
$\kappa$ is a small interaction strength prefactor, $S_z$ 
is a Stokes operator measuring the circularity of optical 
polarization and $F_z$ is the $z$-component of the 
collective atomic spin. Under this Hamiltonian, 
photons with a right circular polarization rotate 
the collective atomic spin over a positive angle 
$\kappa$ along the $z$-axis, while photons with a 
left circular polarization rotate the collective 
spin over a negative angle $-\kappa$. With 
linearly polarized light, the angle of linear 
polarization will Faraday rotate by a degree 
proportional to the $z$-component of the spin.
Note that we entirely neglect `tensor' terms of 
the interaction Hamiltonian (non-linear in individual spin 
operators) which are important near resonance with 
realistic atoms of spin greater than 1/2 \cite{Sto06}.
We have also omitted the evolution due to any driving 
magnetic field, e.g. $H= \gamma B F_y$, purely for 
reasons of simplicity, it can easily be added at 
the end.

In our QSDE-description, the Faraday interaction 
is described as a `direct' scattering process,
without coherent absorption and re-emission.
This is a consequence of the fact that the 
interaction Hamiltonian is derived 
from an approximation in which the excited states 
are adiabatically eliminated \cite{Sto06}. 
At present, however, no mathematically 
rigorous treatment of this elimination 
is available in the literature (see \cite{GvH06} for 
rigorous results on the adiabatic elimination of a 
leaky cavity mode). Therefore, we have chosen 
to directly base our QSDE model on the Faraday 
Hamiltonian without proceeding through a rigorous 
Markov limit \cite{AFLu90, Gou05} followed by 
adiabatic elimination of the excited states. 
Mathematically, the direct scattering is represented 
by gauge-terms in the QSDE \cite{BaL00}. 

Having set the underlying model, i.e.\ the 
QSDE, we rigorously derive the quantum filtering 
equation for the balanced polarimetry setup 
and for homodyne detection of the $y$-polarized 
channel. We investigate the statistics of the 
output processes for these two experiments and take 
a limit where the driving laser power $\alpha^2$ 
goes to infinity but where the product $M=\kappa^2\alpha^2$ 
is kept constant ($\kappa$ is the parameter that couples the field 
to the atomic gas). We show that in this strong driving, weak 
coupling limit the statistics of the output processes for 
the balanced polarimetry experiment 
and the homodyne detection experiment 
are equivalent.  Furthermore, we show that in the strong driving, 
weak coupling limit we obtain the quantum filter that has 
already been intuitively assumed in the literature \cite{Sto06}. 

The remainder of this article is organized as 
follows. Section \ref{sec qsc} introduces the 
fundamental noises and the quantum stochastic 
calculus, and section \ref{sec model} sets our 
QSDE model. Section \ref{sec filter} derives the filter 
when counting in the $45$ degrees rotated $xy$-basis (balanced polarimetry), 
and Section \ref{sec homodyne detection} derives the quantum filter 
for the homodyne detection experiment.
In sections \ref{sec strong driving weak coupling}
and \ref{sec elim} we study the statistics of the observation processes, 
and investigate the strong driving, weak coupling 
limit. We close the paper with a discussion of the 
results obtained.

\section{The quantum calculus}\label{sec qsc}

One polarized photon in a beam of light
can be described by the one particle space 
  \begin{equation*}
  \mathcal{H} = \mathbb{C}^2 \otimes L^2(\mathbb{R}) 
  \cong L^2\big(\mathbb{R};\mathbb{C}^2\big), 
  \end{equation*}
of $\mathbb{C}^2$-valued quadratically integrable 
functions on the real line. The polarized light field is 
described by the \emph{bosonic Fock space} $\mathcal{F}(\mathcal{H})$ over 
$\mathcal{H}$
  \begin{equation*}
  \mathcal{F}(\mathcal{H}) = \mathbb{C} \oplus \bigoplus_{n=1}^\infty
  \mathcal{H}^{\otimes_s n},  
  \end{equation*}
which enables arbitrary superpositions 
between states with a different number of photons. 
Note that photons are bosons and therefore need to 
be described by symmetric wavefunctions. For an 
$f \in \mathcal{H}$ we can define the \emph{exponential 
vector} $e(f)$ in  $\mathcal{F}(\mathcal{H})$ by 
  \begin{equation*}
  e(f) = 1 \oplus \bigoplus_{n=1}^\infty \frac{1}{\sqrt{n!}} f^{\otimes n}.
  \end{equation*}
We call the span of the exponential vectors the 
\emph{exponential domain}. 
The exponential domain is a dense 
set in $\mathcal{F}(\mathcal{H})$ and 
we allow ourselves the freedom to only 
provide the definition of the fundamental noises 
(a little further below) on this domain.   
If we normalize the exponential vectors then 
we obtain the coherent vectors 
$\psi(f) = \exp(-\frac{1}{2} |\!|f|\!|^2)e(f)$.
An important vector is the \emph{vacuum vector}, 
given by $\Phi = \psi(0) = e(0) = 1\oplus 0 \oplus 0 \ldots$.
The \emph{vacuum state} $\phi = \langle\Phi, \,\cdot\ \Phi\rangle$ 
is obtained by taking inner products with the vacuum vector.

If we choose an orthonormal basis $\{e_1,e_2\}$ 
in $\mathbb{C}^2$ then we can decompose every 
$f \in L^2(\mathbb{R};\mathbb{C}^2)$ along 
this basis, i.e.\ $f = f_1 e_1 + f_2 e_2$ with 
$f_1$ and $f_2$ in $L^2(\mathbb{R})$. We now 
introduce the fundamental noises $A^i_t,\ A^{i*}_t$ 
and $\Lambda^{ij}_t$ on the exponential domain 
by (see also \cite{HuP84, Par92, BaL00})
  \begin{equation}\label{eq fundamental}\begin{split}
  &A^i_t e(f) = \left(\int_0^t f_i(s)ds\right)\, e(f),\\
  &\big\langle e(g), A^{i*}_t e(f)\big\rangle  = 
  \left(\int_0^t\overline{g}_i(s)ds\right)\, \big\langle e(g),
  e(f)\big\rangle, \\
  &\big\langle e(g), \Lambda^{ij}_t e(f)\big\rangle  = 
  \left(\int_0^t\overline{g}_i(s)f_j(s)ds\right)\, 
  \big\langle e(g),e(f)\big\rangle. 
  \end{split}\end{equation}
$A^i_t$ and $A^{i*}_t$ are called the 
\emph{annihilation} and \emph{creation}
processes, respectively. The processes 
$\Lambda^{ij}_t$ are called \emph{gauge} 
processes. Formally, we can write the noises 
as $A^i_t = \int_0^t a_s^ids,\ A^{i*}_t = \int_0^t a_s^{i*}ds$ 
and $\Lambda^{ij}_t = \int_0^t a^{i*}_s a^j_s ds$ where 
$a^i_t$ and $a^j_s$ are the usual Bose fields. Mathematically, 
the objects $a^i_t$ and $a^j_s$ are ill-defined and 
therefore we resort to the  
definition of Eq.\ \eqref{eq fundamental}.
The formal expressions do show very explicitly 
though, that the operator $\Lambda^{ii}_t$ counts 
the number of photons with a polarization in the 
$e_i$ direction up to time $t$ and that the 
operator $\Lambda^{ij}_t$ scatters the 
polarization of a photon from the $e_j$ direction 
to the $e_i$ direction.

We will usually work in the basis $\{\vec{e}_x,
\vec{e}_y\}$ which physically corresponds 
to an orthonormal basis in the plane orthogonal to 
the direction of propagation of the light. Apart 
from this basis, we also use the \emph{circular basis} 
given by $\{e_+ = -(\vec{e}_x + 
i\vec{e}_y)/\sqrt{2},\ e_- = 
(\vec{e}_x-i\vec{e}_y)/\sqrt{2}\}$,
and the $45$ degrees rotated $xy$-basis 
given by $\{\vec{e}_\xi = (\vec{e}_x + 
\vec{e}_y)/\sqrt{2},\ \vec{e}_\eta = 
(\vec{e}_x-\vec{e}_y)/\sqrt{2}\}$. 
Given the definitions in Eq.\ \eqref{eq fundamental} it 
is easy to work out how the noises transform under 
basis transformations. For example, we have 
  \begin{equation*}
  \Lambda^{++}_t = \frac{1}{2}\big(\Lambda^{xx}_t + 
  \Lambda^{yy}_t - i\Lambda^{yx}_t + i\Lambda^{xy}_t\big).
  \end{equation*}
  
Denote by $\mathfrak{h}$ the Hilbert space of the 
atomic gas. The space of the combined system of 
atomic gas and field together is then given by  
$\mathfrak{h}\otimes \mathcal{F}(\mathcal{H})$.
Define $\mathcal{H}_{t]} = \mathbb{C}^2\otimes L^2(-\infty,t]$ 
and $\mathcal{H}_{[t} = \mathbb{C}^2\otimes L^2[t,\infty)$. 
For all $t$ the bosonic Fock space splits in a natural
way as a tensor product $\mathcal{F}(\mathcal{H}) = 
\mathcal{F}(\mathcal{H}_{t]})\otimes\mathcal{F}(\mathcal{H}_{[t})$.    
A process $L_s,\, (s\ge 0)$ on $\mathfrak{h}\otimes \mathcal{F}(\mathcal{H})$ 
is called \emph{adapted} if $L_s$ acts nontrivially only 
on $\mathfrak{h}\otimes \mathcal{F}(\mathcal{H}_{s]})$ and 
is the identity on $\mathcal{F}(\mathcal{H}_{[s})$ for 
all $s\ge 0$.

Hudson and Parthasarathy \cite{HuP84} defined stochastic 
integrals of adapted processes $L_s$ against 
the fundamental noises, i.e.\ they gave meaning 
to the expression $X_t = X_0 + \int_0^tL_sdM_s$ where 
$M_s$ is one of the fundamental noises $A^i_s,\ A^{i*}_s$ 
or $\Lambda^{ij}_s$. The expression can be written in 
shorthand as $dX_t = L_tdM_t$. More importantly, 
Hudson and Parthasarathy \cite{HuP84} provided 
the calculus with which these stochastic integrals 
can be manipulated in calculations. The calculus 
consists of the following. Suppose $X_t$ and 
$Y_t$ are stochastic integrals, i.e.\ 
$dX_t = L^1_tdM^1_t$ and $dY_t = L^2_tdM^2_t$ where 
$L^1$ and $L^2$ are adapted processes and $M^1$ 
and $M^2$ are fundamental noises, then the product 
$X_tY_t$ is itself a stochastic integral. Moreover, 
the product $X_tY_t$ satisfies the following 
quantum Ito rule, (partial integration rule)
  \begin{equation*}
  d(X_t Y_t) = X_tdY_t + (dX_t)Y_t + dX_tdY_t, 
  \end{equation*}
where to evaluate $dX_tdY_t$ we use that the 
increment $dM_t$ of a fundamental noise commutes 
with all adapted processes, and products $dM^1_tdM^2_t$ 
are given by the following quantum It\^o table \cite{HuP84}  
\begin{center}
{\large \begin{tabular} {l|lll}
$dM^1\backslash dM^2$ & $dA^{i*}_t$ & $d\Lambda^{ij}_t$ & $dA^i_t$ \\
\hline 
$dA^{k*}_t$ & $0$ & $0$ & $0$ \\
$d\Lambda^{kl}_t$ & $\delta_{li}dA^{k*}_t$ & $\delta_{li}d\Lambda^{kj}_t$ & $0$  \\
$dA^k_t$ & $\delta_{ki}dt$ & $\delta_{ki}dA^j_t$ & $0$ 
\end{tabular} }
\end{center}
and all products $dM_tdt$ and $dtdM_t$ are zero.
As an example, suppose $dX_t = L^1_tdA^i_t$ and 
$dY_t = L^2_tdA^{i*}_t$, then 
$d(X_tY_t) = X_tL^2_tdA^{i*}_t + L^1_tY_tdA^{i}_t + L^1_tL^2_tdt$. 

It can be shown \cite{AFLu90, Gou05} that 
in the weak coupling limit (a Markov limit) 
QED models converge to quantum stochastic models, 
i.e.\ in the limit the unitary time evolution 
$U_t$ satisfies a quantum stochastic differential 
equation in the sense of Hudson and Parthasarathy. 
Usually, the QSDE obtained via the weak coupling 
limit can be simplified further by adiabatic elimination 
of degrees of freedom of the initial system. See 
for instance \cite{GvH06} for rigorous results on the 
adiabatic elimination of a leaky cavity. We, however, 
are interested in the adiabatic elimination of 
the excited states of the atoms in the atomic gas. 
Unfortunately, at present, no rigourous results 
on this kind of adiabatic elimination 
are available. Therefore we choose not to go through a 
weak coupling limit/adiabatic elimination 
procedure here, but rather write down a phenomenological 
QSDE based on the Faraday interaction, 
see Eqs.\ \eqref{eq Ut} and \eqref{eq HuP} 
below. See \cite{Sto06} for a 
derivation of the Faraday Hamiltonian 
of Eq.\ \eqref{eq Faraday Hamiltonian}  
via usual non-rigorous adiabatic 
elimination methods.

\section{The model}\label{sec model}

The interaction between the laser light 
and the spin polarized atomic gas is governed 
by the Faraday interaction given by
  \begin{equation}\label{eq Faraday Hamiltonian}
  Hdt = 2\kappa F_zS_zdt = \kappa F_z 
  \big(d\Lambda^{++}_t - d\Lambda^{--}_t\big). 
  \end{equation}
Here $\kappa$ is a coupling parameter, $F_z$ 
is the $z$-component of the collective 
spin vector of the atoms, and $2S_z = a^{+*}_ta^+_t -a^{-*}_ta^-_t$ is the 
$z$-component of the stokes vector ${\bf S}$ 
of the polarized light. The time evolution of the 
coupled system of light and atomic gas together 
is given by the exponential (the 
superscript $^0$ distinguishes $U_t^0$ from
$U_t$ to be introduced later)
  \begin{equation*}
  U^0_t = \exp\left(i\int_0^t \kappa F_z 
  \big(d\Lambda^{++}_s - d\Lambda^{--}_s\big)\right).
  \end{equation*} 
Since $\Lambda^{++}_t$ and $\Lambda^{--}_t$  
are jump processes, the It\^o rule leads to 
a quantum stochastic differential equation (QSDE) \cite{HuP84}
that contains the following difference terms ($U^0_0 = I$)
  \begin{equation}\label{eq Ut}
  dU^0_t = \left\{\Big(e^{i\kappa F_z} - 1\Big)d\Lambda_t^{++} + 
  \Big(e^{-i\kappa F_z} - 1\Big)d\Lambda_t^{--}\right\}U^0_t.
  \end{equation} 
That is, right circular polarized photons rotate 
the collective spin of the atoms  
over an angle $\kappa$ along the $z$-axis, whereas 
left circular polarized photons rotate the collective 
spin of the atoms over an angle $-\kappa$ 
along the $z$-axis. If we express the gauge 
processes in the linearly polarized $xy$-basis, 
then Eq.\ \eqref{eq Ut} reads ($U^0_0 = I$)
  \begin{equation}\label{eq Ut2}\begin{split}
  dU^0_t = \Big\{\big(\cos(\kappa F_z) -1\big)& \big(d\Lambda_t^{xx}+ d\Lambda_t^{yy}\big)\ -\\
  &\sin(\kappa F_z)\big(d\Lambda_t^{xy} - d\Lambda^{yx}_t\big)\Big\}U^0_t.
  \end{split}\end{equation}
The second term shows that the interaction can
scatter $x$-polarized photons to $y$-polarized photons and 
vice versa.

\begin{figure}
\begin{center}
\includegraphics[width=3in]{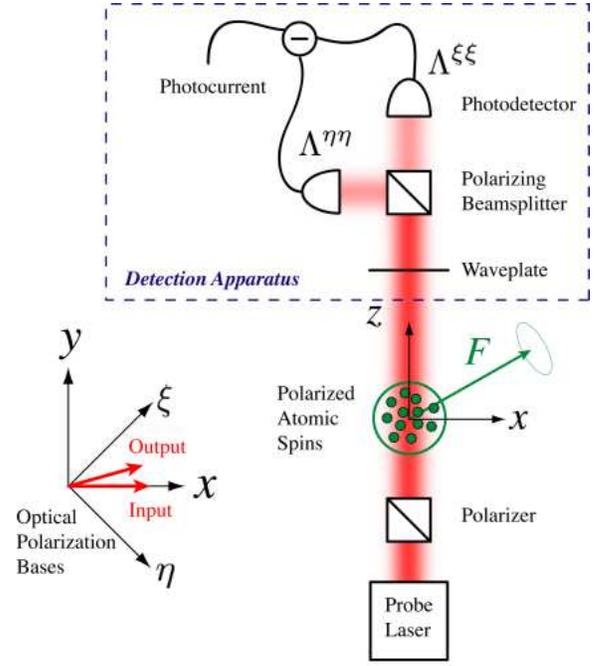}
\caption[PolarimeterLuc]{\label{Fig one} Schematic depicting balanced polarimetric  
detection of laser light after interacting with a polarized cloud of  
atomic spins via the Faraday Hamiltonian.  The light is initially  
linearly polarized along the x direction.  After the interaction, the  
light carries off information about the atomic gas encoded in a small  
optical polarization rotation.  The light is measured in the $\xi$-$ 
\eta$ basis rotated $45$ degrees from the $x$-$y$ basis, such that  
without the atomic gas the mean output of the polarimeter is balanced  
to zero.  The change of measurement basis is achieved with the  
waveplate located just before the polarizing beamsplitter.}
\end{center}
\end{figure}

Initially, the atomic gas is in an $x$-spin polarized
state, denoted $\rho$, and the field is in an $x$-polarized 
coherent state $\psi^x(f)$ which represents the driving 
laser. The function $f \in L^2(\mathbb{R}^+)$ gives
the phase and amplitude of the driving laser field at 
every time $t \in \mathbb{R}^+$. In computations it is 
often convenient to work with respect to the vacuum state 
$\phi = \langle\Phi,\, \cdot\, \Phi\rangle$ for the field. 
We can obtain a coherent state by acting with a 
\emph{displacement} or \emph{Weyl operator} $W^x(f)$ 
on the vacuum vector
  \begin{equation*}
  \psi^x(f) = W^x(f)\Phi. 
  \end{equation*}  
If we work with respect to the vacuum state $\phi$, then 
we have to sandwich all operators with the Weyl 
operator $W^x(f)$. An observable $S$ of the combined system 
of atomic gas and field up to time $t$ is therefore at time 
$t$ given by
  \begin{equation}\label{eq split at t}\begin{split}
  j_t(S) ={ } &W^x(f)^*U^{0*}_tSU_t^0W^x(f) \\ 
  ={ } &W^x(f_{t})^*U_t^{0*}SU_t^0W^x(f_{t}).   
  \end{split}\end{equation}
Here, $f_t$ denotes the function $f$ truncated at time 
$t$, i.e.\ $f_t(s) = f(s)$ for all $s\le t$ and 
$f_t(s) = 0$ for all $s >t$. The relation Eq.\ \eqref{eq split at t}
follows since all the operators split as a tensor product at 
time $t$ and $U^0_t$ and $S$ act as the identity operator 
after time $t$. Since $W^x(f)$ is unitary, it then cancels against 
its adjoint for the part that is after time $t$.

It can be shown \cite{Par92} that $W^x(f_t)$ satisfies 
the following QSDE ($W^x(f_0) = I$)
  \begin{equation*}
  dW^x(f_t) = \left\{f(t)dA^{x*}_t - \overline{f}(t)dA^x_t - \frac{1}{2}|f(t)|dt\right\}W^x(f_t).
  \end{equation*}
Defining $U_t = U^{0}_t W^x(f_t)$ and using the quantum It\^o rule \cite{HuP84}, 
we obtain ($U_0 = I$)
  \begin{equation}\label{eq HuP}\begin{split}
  &dU_t = \Big\{\big(\cos(\kappa F_z) -1\big) \big(d\Lambda_t^{xx}+ d\Lambda_t^{yy}\big)\ -\\
  &\sin(\kappa F_z)\big(d\Lambda_t^{xy} -  d\Lambda^{yx}_t\big) + 
  f(t)\cos(\kappa F_z)dA^{x*}_t\ - \\ 
  &\overline{f}(t)dA^x_t + f(t)\sin(\kappa F_z)dA_t^{y*} - \frac{1}{2}|f(t)|^2dt  
  \Big\}U_t.
  \end{split}\end{equation}
Summarizing, we work in the state $\mathbb{P} := \rho\otimes\phi$, 
the time evolution of (adapted) observables $S$ is 
given by $j_t(S) = U^*_tSU_t$,  
with $U_t$ given by Eq.\ \eqref{eq HuP}.

\section{The quantum filter}\label{sec filter}

After the interaction, 
the light carries off information about
the atomic gas. Therefore, measuring the field 
will enable us to make inference about the 
atomic gas observables. Let us suppose that we are 
counting the photons with a polarization 
along the $\vec{e}_\xi = 1/\sqrt{2}
(\vec{e}_x + \vec{e}_y)$ 
axis, and that we are separately counting 
the photons with a polarization along the 
$\vec{e}_\eta = 
1/\sqrt{2}(\vec{e}_x  - \vec{e}_y)$
axis, see Fig.\ \ref{Fig one}. That is, our observations are given by 
  \begin{equation}\label{eq obs}\begin{split}
  &Y^\xi_t = U_t^*\Lambda^{\xi\xi}_tU_t = 
  \frac{1}{2}U_t^*\big(\Lambda_t^{xx}+ \Lambda_t^{yy}+\Lambda_t^{xy}+\Lambda_t^{yx}\big)U_t,\\
  &Y^\eta_t = U_t^*\Lambda^{\eta\eta}_tU_t = 
  \frac{1}{2}U_t^*\big(\Lambda_t^{xx}+ \Lambda_t^{yy}-\Lambda_t^{xy}-\Lambda_t^{yx}\big)U_t. 
  \end{split}\end{equation}  
Let $X$ be an atomic gas operator, its time 
evolution is given by 
  \begin{equation}\label{eq system}
  j_t(X) = U^*_tXU_t.
  \end{equation}
Eq.\ \eqref{eq system} is called \emph{the system}. 
Together Eqs.\ \eqref{eq system} and \eqref{eq obs} 
form a \emph{system-observations pair}.

It is easily checked that $[Y^\alpha_t, Y^\beta_s] =0$ for all 
$\alpha,\beta \in\{\xi,\eta\}$ and for all $t,s \ge 0$. This is 
called the \emph{self-nondemolition} property and ensures that 
our observations are simultaneously observable classical 
processes. Furthermore, it can be shown that 
$[j_t(X), Y_s^\alpha] = 0$ for all $t \ge s \ge 0$ and 
$\alpha \in \{\xi,\eta\}$. This is called the \emph{nondemolition} 
property. Together the self-nondemolition and the nondemolition 
property ensure the existence of the conditional expectation 
$\mathbb{P}(j_t(X)|\mathcal{Y}_t)$ of a system operator 
at time $t$ on the observations up to time $t$. Since 
the conditional expectation is linear in the atomic gas 
operators $X$, we can define an \emph{information state} 
$\pi_t$ on the atomic gas system by 
  \begin{equation*} 
  \pi_t(X) = \mathbb{P}(j_t(X)|\mathcal{Y}_t).
  \end{equation*}
Note that $\pi_t$ is a stochastic state since it depends 
on the observations $Y^\xi$ and $Y^\eta$ up to time $t$.
  
It is the goal of quantum filtering theory to obtain a recursive 
stochastic differential equation that propagates the 
information state $\pi_t$ in time. Our approach here is 
based on the reference probability method \cite{BVH05a,BVH05b}. 
In these references, the interested reader can find further 
details on the exposition below. 

Our first step is one of mere convenience. It is a change 
of picture that will simplify subsequent calculations. 
Let $W_t$ be given by $W_0 = I$ and
  \begin{equation*}
  dW_t =  \left\{\overline{f}(t)dA^{x}_t - f(t)dA^{x*}_t - \frac{1}{2}|f(t)|dt\right\}W_t.
  \end{equation*}
Note that $W_t$ is the adjoint of $W^x(f_t)$. Now define 
$U_t' = W_tU_t$, where $U_t$ is given by Eq.\ \eqref{eq HuP}.
It easily follows from the quantum It\^o rule \cite{HuP84} 
that 
  \begin{equation}\label{eq Utprime}\begin{split}
  &dU_t' = \Big\{\big(\cos(\kappa F_z) -1\big) \big(d\Lambda_t^{xx}+ d\Lambda_t^{yy}\big)\ -\\
  &\sin(\kappa F_z)\big(d\Lambda_t^{xy} -  d\Lambda^{yx}_t\big) + 
  f(t)\big(\cos(\kappa F_z)-1\big)dA^{x*}_t\ - \\ 
  &\overline{f}(t)\big(\cos(\kappa F_z) -1\big)dA^x_t + 
  f(t)\sin(\kappa F_z)dA_t^{y*}\ - \\ 
  &\overline{f}(t)\sin(\kappa F_z)dA^y_t + |f(t)|^2\big(\cos(\kappa F_z)-1\big)dt  
  \Big\}U_t'.
  \end{split}\end{equation}
Define a 
new state on the combined system of atomic gas and field 
by $\mathbb{Q}^t(S) = \mathbb{P}({U_t'}^*SU_t')$.
To complete our change of picture we need to sandwich 
the observables with the opposite rotation. That means 
that  the system Eq.\ \eqref{eq system} is now simply 
given by $U_t'U^*_tXU_t{U_t'}^* = W_tXW_t^* = X$. In 
the last step we used that $X$ acts on the atoms 
and is the identity on the field and 
vice versa for $W_t$. In the new picture the observations 
read
  \begin{equation*}\begin{split}
  &Z_t^{\xi} = U_t'Y_t^{\xi}{U_t'}^* = W_t\Lambda_t^{\xi\xi}W_t^*,\\
  &Z_t^{\eta} =  U_t'Y_t^{\eta}{U_t'}^* = W_t\Lambda_t^{\eta\eta}W_t^*.  
  \end{split}\end{equation*}
Using the quantum It\^o rule, it easily follows that
  \begin{equation}\begin{split}
  &dZ_t^{\xi} = d\Lambda^{\xi\xi}_t + \frac{1}{2}
  \big(\overline{f}(t)(dA^x_t + dA^y_t)\ + \\
  &\ \ \ \ \ \ \ \ \ \ \ \ \ \ \ \ f(t)(dA^{x*}_t + dA^{y*}_t) + |f(t)|^2dt\big),\\
  &dZ_t^{\eta} = d\Lambda^{\eta\eta}_t + \frac{1}{2}
  \big(\overline{f}(t)(dA^x_t - dA^y_t)\ + \\
  &\ \ \ \ \ \ \ \ \ \ \ \ \ \ \ \ f(t)(dA^{x*}_t - dA^{y*}_t) + |f(t)|^2dt\big).
  \end{split}\end{equation}
Denote $\mathcal{C}_t = U_t'\mathcal{Y}_t{U_t'}^*$, i.e.\ $\mathcal{C}_t$
consists of the 
processes $Z^{\xi}$ and $Z^{\eta}$ up to time $t$.
It can easily be shown that the conditional expectations 
in the two different pictures are related by 
$\mathbb{P}(j_t(X)|\mathcal{Y}_t) = 
{U_t'}^*\mathbb{Q}^t(X|\mathcal{C}_t)U_t'$.  
That completes our discussion of the change of 
picture. We will now focus on deriving an equation 
that propagates $\mathbb{Q}^t(X|\mathcal{C}_t)$.

At the heart of the reference probability method 
is the following \emph{quantum Bayes formula} 
\cite{BVH05a,BVH05b}. Let $V$ be an operator 
that commutes with $Z^\alpha_s$ for all 
$\alpha \in \{\xi,\eta\}$ and for all $0 \le s\le t$.
Moreover, suppose that $V^*V > 0$ and that 
$\mathbb{P}(V^*V) =1$. Then we can define a 
state $\mathbb{Q}$ by 
$\mathbb{Q}(S) = \mathbb{P}(V^*SV)$, and for 
all operators $X$ that commute with 
$Z^\alpha_s\ (\alpha \in \{\xi,\eta\}, 0\le s\le t)$, 
we have (see \cite{BVH05a,BVH05b} for a proof)
  \begin{equation*}
  \mathbb{Q}(X|\mathcal{C}_t) = 
  \frac{\mathbb{P}(V^*XV|\mathcal{C}_t)}{\mathbb{P}(V^*V|\mathcal{C}_t)}.
  \end{equation*} 
We would like to apply the quantum Bayes formula 
to $\mathbb{Q}^t$, i.e.\ with $V = U_t'$. However, Eq.\ \eqref{eq Utprime}
shows that $U_t'$ is driven by noises that do not commute with 
$Z^\alpha_s\ (\alpha \in \{\xi,\eta\}, 0\le s\le t)$, i.e.\ 
$U_t'$ itself does not commute with the $Z^\alpha_s$'s.

The following trick \cite{Hol90} solves this problem. 
Suppose $V_t'$ satisfies the QSDE
  \begin{equation}\label{eq Vprime}\begin{split}
  dV_t' = \Big\{\big(\cos(\kappa F_z)& + \sin(\kappa F_z)-1\big)dZ_t^\xi\ + \\
  &\big(\cos(\kappa F_z) - \sin(\kappa F_z)-1\big)dZ_t^\eta\Big\}V_t'.
  \end{split}\end{equation}  
Then, the coefficients of $dA_t^{x*}$, $dA_t^{y*}$
and $dt$ are the same as in Eq.\ \eqref{eq Utprime}. Since $dA^\alpha_t$ 
and $d\Lambda_t^{\alpha\beta}$ ($\alpha,\beta \in \{x,y\}$) 
are zero when acting on the vacuum vector $\Phi$, we therefore 
have that for all operators $S$ \cite{Hol90}
  \begin{equation*}
  \mathbb{Q}^t(S) = \mathbb{P}\big({U_t'}^*SU_t'\big) = \mathbb{P}({V_t'}^*SV_t'). 
  \end{equation*}
Moreover, since $V_t'$ is driven by $Z_t^\xi$ and $Z_t^\eta$, it commutes 
with $\mathcal{C}_t$, and we can therefore apply the Bayes 
formula with $V = V_t'$. That is, summarizing what we have achieved 
thus far
  \begin{equation}\label{eq summary}
  \mathbb{P}\big(j_t(X)|\mathcal{Y}_t\big) = {U_t'}^*\mathbb{Q}^t(X|\mathcal{C}_t)U_t' =
  \frac{{U_t'}^*\mathbb{P}\big({V_t'}^*XV_t'|\mathcal{C}_t\big)U_t'}
  {{U_t'}^*\mathbb{P}\big({V_t'}^*V_t'|\mathcal{C}_t\big)U_t'}. 
  \end{equation}  
The next step is to find the equation that 
propagates $\mathbb{P}({V_t'}^*XV_t'|\mathcal{C}_t)$ 
in time. Using the quantum It\^o rule we find
  \begin{equation}\label{eq VXV}\begin{split} 
  &d{V_t'}^*XV_t' = {V_t'}^*\big(L^\xi XL^\xi -X\big)V_t'dZ_t^\xi\ + \\
  &\ \ \ \ \ \ \ \ \ \ \ \ \ \ \ \ \ \ \ \ \ \ \ \ \
  {V_t'}^*\big(L^\eta XL^\eta -X\big)V_t'dZ_t^\eta, 
  \end{split}\end{equation}
with 
  \begin{equation}\label{eq LxiLeta}\begin{split}
  &L^\xi = \cos(\kappa F_z) +\sin(\kappa F_z), \\  
  &L^\eta = \cos(\kappa F_z) -\sin(\kappa F_z). 
  \end{split}\end{equation}
We can write Eq.\ \eqref{eq VXV} in integral form and approximate 
the stochastic integrals in the usual way with simple 
processes. If we proceed by taking the conditional expectation 
$\mathbb{P}(\,\cdot\,|\mathcal{C}_t)$, then we can 
pull the integrators which are elements of $\mathcal{C}_t$
out of the expectation. Furthermore, the conditional 
expectation $\mathbb{P}(L_s|\mathcal{C}_t), (0\le s\le t)$ of 
an adapted process $L$ equals $\mathbb{P}(L_s|\mathcal{C}_s)$.    
In this way we obtain
  \begin{equation*}\begin{split}
  &d\mathbb{P}\big({V_t'}^*XV_t'\big|\mathcal{C}_t\big) =
  \mathbb{P}\big({V_t'}^*\big(L^\xi XL^\xi -X\big)V_t'\big|\mathcal{C}_t\big)dZ_t^\xi\ + \\
  &\ \ \ \ \ \ \ \ \ \ \ \ \ \ \ \ \ 
  \mathbb{P}\big({V_t'}^*\big(L^\eta XL^\eta -X\big)V_t'\big|\mathcal{C}_t\big)dZ_t^\eta.
  \end{split}\end{equation*}
Now define $\sigma_t(X) = {U_t'}^*\mathbb{P}({V_t'}^*XV_t'|\mathcal{C}_t)U_t'$ 
for all atomic operators $X$. Using the quantum 
It\^o rule, we obtain the linear version 
of the quantum filtering equation
  \begin{equation}\label{eq Zakai}\begin{split}
  &d\sigma_t(X) = \sigma_t\big(\mathcal{L}(X)\big)dt\ + \\
  &\sigma_t\big(L^\xi XL^\xi -X\big)\left(dY_t^\xi - \frac{1}{2}|f(t)|^2dt\right)\  + \\
  &\sigma_t\big(L^\eta X L^\eta -X\big)\left(dY_t^\eta - \frac{1}{2}|f(t)|^2dt\right),
  \end{split}\end{equation}
where the Lindblad generator $\mathcal{L}$ is given by
  \begin{equation}\label{eq Lindblad}\begin{split}
  \mathcal{L}(X) = |f(t)|^2\Big(&\sin(\kappa F_z) X\sin(\kappa F_z)\ + \\ 
  &\cos(\kappa F_z) X\cos(\kappa F_z) -X\Big),
  \end{split}\end{equation}
for all atomic operators $X$. Now recall from 
Eq.\ \eqref{eq summary} that 
$\pi_t(X) = \sigma_t(X)/\sigma_t(I)$, which 
is a quantum version of the classical \emph{Kallianpur-Striebel} 
formula. Using the It\^o rule once more, we obtain 
the following quantum filter
  \begin{equation*}\begin{split}
  &\ \ \ \ \ \ \ \ \ \ \ d\pi_t(X) = \pi_t\big(\mathcal{L}(X)\big) dt\ + \\ 
  &\left(\frac{\pi_t\big(L^\xi XL^\xi\big)}{\pi_t\big(L^\xi L^\xi\big)} 
  -\pi_t(X)\right)\left(dY_t^\xi - \frac{1}{2}|f(t)|^2\pi_t\big(L^{\xi}L^\xi\big)dt\right)  + \\
  &\left(\frac{\pi_t\big(L^\eta X L^\eta\big)}{\pi_t\big(L^\eta L^\eta\big)} 
  - \pi_t(X)\right)\left(dY_t^\eta - \frac{1}{2}|f(t)|^2\pi_t\big(L^\eta L^\eta\big)dt\right).
  \end{split}\end{equation*}
The processes  
$dY_t^\xi-  \frac{1}{2}|f(t)|^2\pi_t\big(L^{\xi}L^\xi\big)dt$ and 
$dY_t^\eta - \frac{1}{2}|f(t)|^2\pi_t\big(L^\eta L^\eta\big)dt$ 
are called the \emph{innovations} or \emph{innovating martingales}. 
It can indeed be shown \cite{Bel92b, BGM04} that the 
innovations are martingales with respect 
to the filtration $\mathcal{Y}_t$ and the measure induced by 
$\mathbb{P}$.

\section{A different setup: homodyne detection}\label{sec homodyne detection}

\begin{figure}
\begin{center}
\includegraphics[width=3in]{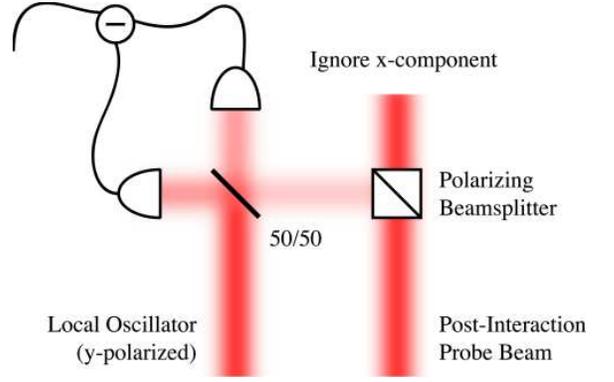}
\caption[Homodyneluc]{\label{Fig two} In the homodyne detection setup, 
the detection apparatus in the dashed box of Fig.\ \ref{Fig one} 
should be replaced with the apparatus depicted schematically here.  
As in Fig.\ \ref{Fig one}, the light is initially polarized along x  
and the polarization rotates slightly due to the interaction with the  
atoms.  Here, however, the strong x component is split off and  
ignored while the weak y component is sent to a standard homodyne 
setup.  The y component is mixed at a 50/50 (non-polarizing)  
beamsplitter with a strong local oscillator beam also polarized along  
the y direction and derived from the same laser as the probe beam.   
The photocurrent representing the interference signal is then derived  
from the difference between the outputs of the photodetectors.}
\end{center}
\end{figure}

For the remainder of the paper 
we assume that $f(t) = \alpha e^{i\phi_t}$ 
with $\alpha$ real and $\phi_t$ 
in $[0,2\pi)$. Now suppose that  
instead of the balanced polarimetry 
setup described in the previous section, 
we use a homodyne detection setup 
to measure the $y$-component of 
the output light, see Fig.\ \ref{Fig two}. It is well known 
\cite{Bar90,Car93} that for such a homodyne 
detection setup the observations 
are given by 
  \begin{equation}\label{eq obs homodyne}
  Y_t = U_t^*(e^{-i\phi_t}A^y_t + e^{i\phi_t}A^{y*}_t)U_t.
  \end{equation}
That is, for homodyne detection of 
the $y$-channel the system-observations pair 
is given by Eqs.\ \eqref{eq system} 
and \eqref{eq obs homodyne}. It is 
easily checked that the homodyne system-observations 
pair satisfies the self-nondemolition 
and nondemolition properties, meaning 
that the conditional expectation 
$\mathbb{P}(j_t(X)|\mathcal{Y}_t)$ is well-defined. 
Here $\mathcal{Y}_t$ denotes the homodyne 
observations of Eq.\ \eqref{eq obs homodyne} from 
time $0$ up to time $t$. We will now 
derive the filter for the corresponding 
information state $\pi_t(X) = \mathbb{P}(j_t(X)|\mathcal{Y}_t)$.  

Our first step is again one of convenience. 
We change to the Schr\"odinger picture by 
defining the following state on 
the combined system of atomic gas and 
field together $\mathbb{Q}^t(S) = \mathbb{P}(U_t^*SU_t)$. 
In the Schr\"odinger picture our system is simply given 
by $U_tj_t(X)U_t^* = X$ and the observations 
are given by 
  \begin{equation*}
  Z_t = U_tY_tU_t^* = e^{-i\phi_t}A^y_t + e^{i\phi_t}A^{y*}_t. 
  \end{equation*}
Denote $\mathcal{C}_t = U_t\mathcal{Y}_tU_t^*$, i.e.\ 
$\mathcal{C}_t$ consists of the process $Z$ up 
to time $t$. It can easily be shown that the conditional 
expectations in the Heisenberg and Schr\"odinger pictures 
are related by $\mathbb{P}(j_t(X)|\mathcal{Y}_t) = 
U_t^*\mathbb{Q}^t(X|\mathcal{C}_t)U_t$.

To compute $\mathbb{Q}^t(X|\mathcal{C}_t)$ we 
would like to use the Bayes formula. 
Suppose $V_t$ satisfies the following 
QSDE ($V_0 =I$)
  \begin{equation}\label{eq V}
  dV_t = \Big\{e^{i\phi_t}\alpha\cos(\kappa F_z)dA^{x*}_t +
  \alpha \sin(\kappa F_z)dZ_t 
  - \frac{\alpha^2}{2}dt  
  \Big\}V_t. 
  \end{equation} 
Then, the coefficients of $dA^{x*}_t,\ dA_t^{y*}$ 
and $dt$ are the same as in Eq.\ \eqref{eq HuP}.
Therefore we have $\mathbb{Q}^t(S) = \mathbb{P}(U_t^*SU_t) = 
\mathbb{P}(V_t^*SV_t)$. The equation for 
$V_t$ is driven by $Z$ and $A^{x*}$, 
both commute with $\mathcal{C}_t$, i.e.\ 
$V_t$ commutes with $\mathcal{C}_t$. That 
means we can now apply Bayes formula with 
$V = V_t$ to obtain
  \begin{equation*}
    \mathbb{P}\big(j_t(X)|\mathcal{Y}_t\big) = 
    {U_t}^*\mathbb{Q}^t(X|\mathcal{C}_t)U_t =
  \frac{{U_t}^*\mathbb{P}\big({V_t}^*XV_t|\mathcal{C}_t\big)U_t}
  {{U_t}^*\mathbb{P}\big({V_t}^*V_t|\mathcal{C}_t\big)U_t}.
  \end{equation*}

Using the quantum It\^o rule we find
  \begin{equation*}\begin{split}
  dV_t^*XV_t ={ } &V_t^*\mathcal{L}(X)V_tdt\ + \\ 
  &\alpha e^{-i\phi_t} V_t^*\cos(\kappa F_z)XV_t dA_t^x\ + \\ 
  &\alpha e^{i\phi_t} V_t^*X\cos(\kappa F_z)V_tdA_t^{x*}\ + \\
  &\alpha V_t^*\big(\sin(\kappa F_z)X + X\sin(\kappa F_z)\big)V_tdZ_t.
  \end{split}\end{equation*}
where $\mathcal{L}$ is given by Eq.\ 
\eqref{eq Lindblad}. Since $dA_t^x$ and 
$dA^{x*}_t$ are independent of $\mathcal{C}_t$ 
and since vacuum expectations of stochastic 
integrals with respect to $dA_t^x$ and 
$dA_t^{x*}$ are zero, we find in an analogous 
way as before
  \begin{equation*}
  \begin{split}
  &d\mathbb{P}\big(V_t^*XV_t\big|\mathcal{C}_t\big) 
  = \mathbb{P}\big(V_t^*\mathcal{L}(X)V_t\big|\mathcal{C}_t\big)dt\ + \\ 
  &\alpha \mathbb{P}\Big(V_t^*\big(\sin(\kappa F_z)X + 
  X\sin(\kappa F_z)\big)V_t\Big|\mathcal{C}_t\Big)dZ_t.
  \end{split}\end{equation*}
Now introduce $\sigma_t(X) = U_t^*\mathbb{P}(V_t^*XV_t|\mathcal{C}_t)U_t$ 
for all atomic gas operators $X$. Using the 
quantum It\^o rule, we obtain 
the linear homodyne filtering equation 
  \begin{equation}\label{eq homodyne Zakai}\begin{split}
  d\sigma_t(X) ={ } &\sigma_t\big(\mathcal{L}(X)\big)dt\ + \\ 
  &\alpha\sigma_t\big(\sin(\kappa F_z)X + 
  X\sin(\kappa F_z)\big)dY_t.
  \end{split}\end{equation}
Using $\pi_t(X) = \sigma_t(X)/\sigma_t(I)$ 
and the It\^o rule, we find the following 
homodyne quantum filter 
  \begin{equation}\label{eq homodyne filter}\begin{split}
  &d\pi_t(X) = \pi_t\big(\mathcal{L}(X)\big)dt\ + \\ 
  &\alpha\Big(\pi_t\big(\sin(\kappa F_z)X + 
  X\sin(\kappa F_z)\big)- 2\pi_t\big(\sin(\kappa F_z)\big)
  \pi_t(X)\Big)\\
  &\times\Big(dY_t-2\alpha\pi_t\big(\sin(\kappa F_z)\big)dt\Big).
  \end{split}\end{equation}
The process 
$dY_t-2\alpha\pi_t\big(\sin(\kappa F_z)\big)dt$
is again called the innovations or the innovating 
martingale. It can be shown \cite{Bel92b,BGM04} 
that the innovations are a continuous martingale 
with respect to the filtration $\mathcal{Y}_t$ and the measure 
induced by $\mathbb{P}$. It follows from Levy's 
theorem that the innovations for the homodyne 
detection setup form a Wiener process.

\section{Strong driving, weak coupling}\label{sec strong driving weak coupling}

Define the \emph{measurement strength} as 
the product $M = \alpha^2\kappa^2$.
In a typical experimental setting $\alpha$ 
will be very large (strong driving) and 
$\kappa$ will be very small (weak coupling). 
The idea in this section will be to exaggerate 
this by taking the limit $\alpha \to 
\infty$ while keeping the product $M=\alpha^2\kappa^2$ 
constant. 

Let us introduce the following scaled sum and 
difference processes 
  \begin{equation}\label{eq scaled sum difference}
  Y^+_t = \frac{Y^\xi_t + Y^\eta_t}{\alpha^2}, \ \ \ \ 
  Y^-_t = \frac{Y^\xi_t - Y^\eta_t}{\alpha}.
  \end{equation}
Note that we scaled the sum by $\alpha^2$ and 
the difference by $\alpha$. We will see that with 
these scalings we get finite output processes in 
the limit. In practice the scalings are determined by 
the experiment, i.e.\ they are chosen in such a way 
that the photocurrents nicely fill the scales on the 
read out devices. 
We are interested in the statistics of the 
processes $Y_t^+$ and $Y_t^-$. Therefore, following 
\cite{Bar90}, we introduce their 
characteristic functionals
  \begin{equation*}\begin{split}
  \Phi^+(k,t) ={ } &\mathbb{P}\left(\exp\left(-i\int_0^tk(s)dY^+_s\right)\right) \\
  ={ } &\mathbb{P}\left(U_t^*\exp\left(-i\int_0^t\frac{k(s)}{\alpha^2}
  (d\Lambda_s^{xx}+d\Lambda_s^{yy})\right)U_t\right),\\
  \Phi^-(k,t) ={ } &\mathbb{P}\left(\exp\left(-i\int_0^tk(s)dY^-_s\right)\right) \\
  ={ } &\mathbb{P}\left(U_t^*\exp\left(-i\int_0^t\frac{k(s)}{\alpha}
  (d\Lambda_s^{xy}+d\Lambda_s^{yx})\right)U_t\right), 
  \end{split}\end{equation*} 
where $k$ is an arbitrary function in $L^2(\mathbb{R}^+)$.
The characteristic functionals $\Phi^+$ and $\Phi^-$ 
faithfully encode the complete statistics of the 
processes $Y_t^+$ and $Y^-_t$. 

Using the quantum It\^o rule and the fact that vacuum 
expectations of stochastic integrals are zero, we 
find the following differential equation for $\Phi^+(k,t)$
  \begin{equation*}
  \frac{d\Phi^+}{dt}(k,t) = \alpha^2\left(\exp\left(-i\frac{k(t)}{\alpha^2}\right)
  -1\right)\Phi^+(k,t),
  \end{equation*}
In the limit $\alpha$ to infinity (while keeping $M=\alpha^2\kappa^2$ constant), 
we therefore obtain
  \begin{equation*}
  \Phi^+(k,t) = \exp\left(-i\int_0^tk(s)ds\right).
  \end{equation*}
This is the characteristic functional of the deterministic 
time process $t$. In short, as $\alpha$ tends to infinity, 
$dY_t^+$ tends to $dt$.

To calculate $\Phi^-(k,t)$, define 
for all atomic gas operators $X$
  \begin{equation*}\begin{split}
  &\Phi^-(X,k,t) = \\ 
  &\mathbb{P}\left(U_t^*X\exp\left(-i\int_0^t\frac{k(s)}{\alpha}
  (d\Lambda_s^{xy}+d\Lambda_s^{yx})\right)U_t\right).
  \end{split}\end{equation*}
Note that $\Phi^-(k,s) = \Phi^-(I,k,s)$.   
Using the quantum It\^o rule and the fact that vacuum 
expectations of stochastic integrals are zero, we 
find the following system of differential 
equations ($n\ge 0$)
  \begin{equation*}\begin{split}
  &\frac{d\Phi^-}{dt}\Big(\alpha^n\sin^n(2\kappa F_z),k,t\Big) =  \\
  &\alpha^2 \Big(\cos\left(\frac{k(t)}{\alpha}\right)-1\Big)
  \Phi^-\Big(\alpha^n\sin^n(2\kappa F_z),k,t\Big)\ - \\ 
  &i \alpha\sin\left(\frac{k(t)}{\alpha}\right)\Phi^-\Big(\alpha^{n+1}\sin^{n+1}(2\kappa F_z),k,t\Big).
  \end{split}\end{equation*}
Note that although the atomic gas system might be 
very high dimensional, the dimension is finite. That 
means that the above system of differential equation 
is closed and consists only of a finite number of equations.   
In the limit $\alpha$ to infinity 
(while keeping $M=\alpha^2\kappa^2$ constant), 
we obtain the following finite system of coupled 
differential equations ($n\ge 0$)  
  \begin{equation}\label{eq difeqs}
  \begin{split}
  &\frac{d\Phi^-}{dt}\Big((\sqrt{M}F_z)^n,k,t\Big) =  
  -\frac{k(t)^2}{2}\Phi^-\Big((\sqrt{M}F_z)^n,k,t\Big)\ - \\ 
  &\ \ \ \ \ \ \ \ \ \ \ \ \ \ 2ik(t)\Phi^-\Big((\sqrt{M}F_z)^{n+1},k,t\Big).
  \end{split}
  \end{equation} 
In principle we could now try to solve this system 
of equations. 
However, instead of finding an explicit solution, 
let us compare this with the statistics of the 
homodyne observations $Y_t$ 
defined in Eq.\ \eqref{eq obs homodyne}. In 
analogy to the discussion above, we define 
for all atomic gas operators $X$
  \begin{equation*}\begin{split}
  &\Phi(X,k,t) =  \\ 
  &\mathbb{P}\left(U_t^*X\exp\left(-i\int_0^t k(s)
  d(e^{-i\phi_t}A_s^y + e^{i\phi_t}A_s^{y*})\right)U_t\right).
  \end{split}\end{equation*}
Using the quantum It\^o rule and the fact that vacuum 
expectations of stochastic integrals are zero, we 
find the following system of differential 
equations ($n\ge 0$)
  \begin{equation*}\begin{split}
  \frac{d\Phi}{dt}\Big(\alpha^n&\sin^n(\kappa F_z),k,t\Big) = \\
  &-\frac{k(t)^2}{2}\Phi\Big(\alpha^n\sin^n(\kappa F_z),k,t\Big)\ - \\
  &2ik(t)\Phi\Big(\alpha^{n+1}\sin^{n+1}(\kappa F_z),k,t\Big).
  \end{split}\end{equation*}
Taking the limit $\alpha \to \infty$ (while $M=\alpha^2\kappa^2$ is 
held constant) then again leads to the system of differential 
equations Eq.\ \eqref{eq difeqs}. Therefore we conclude 
that in the limit the processes $Y^-_t$ and 
$Y_t$ have exactly the 
same statistics! 
This means that from the point of view of 
statistical inference of the atomic gas system 
from the observations, the balanced polarimetry 
experiment and the $y$-channel 
homodyne detection experiment 
are equivalent.

Rearranging terms, we can write the 
linear quantum filtering equation Eq.\ \eqref{eq Zakai} 
as 
  \begin{equation*}\begin{split}
  d\sigma_t(X) ={ }&\alpha^2\Bigg(\sigma_t\big(\sin(\kappa F_z)X\sin(\kappa F_z)\big)\ + \\
  &\sigma_t\big(\cos(\kappa F_z)X \cos(\kappa F_z)\big)-\sigma_t(X)\Bigg) dY_t^+\ + \\
  &\alpha\Bigg(\sigma_t\big(\cos(\kappa F_z)X\sin(\kappa F_z)\big)\ + \\  
  &\sigma_t\big(\cos(\kappa F_z)X \sin(\kappa F_z)\big)\Bigg)dY_t^-.
  \end{split}\end{equation*} 
Writing $\overline{Y}_t$ for the limit process of $Y_t^-$, and 
taking the limit of the above equation, we obtain the 
following linear quantum filtering equation
  \begin{equation}\label{eq limit linear filter}
  d\sigma_t(X) = \sigma_t\big(\overline{\mathcal{L}}(X)\big)dt\ + \
  \sqrt{M}\sigma_t\big(F_zX + XF_z\big)d\overline{Y}_t, 
  \end{equation}
where 
  \begin{equation*}
  \overline{\mathcal{L}}(X) = M\Big(F_zXF_z -\frac{1}{2}\big(F_z^2 X + X
  F^2_z\big)\Big). 
  \end{equation*}  
Moreover, we obtain the following 
normalized quantum filter 
  \begin{equation}\label{eq limit filter}\begin{split}
  &d\pi_t(X) = \pi_t\big(\overline{\mathcal{L}}(X)\big)dt +
  \sqrt{M}\Big(\pi_t(F_z X + XF_z)\ -\\ 
  &2\pi_t(F_z)\pi_t(X)\Big)
  \Big(d\overline{Y}_t - 2\sqrt{M}\pi_t(F_z)dt\Big).
  \end{split}\end{equation}
Since $d\overline{Y}_t-2\sqrt{M}\pi_t(F_z)dt$ is a 
continuous martingale \cite{Bel92b, BGM04}, 
it follows from Levy's theorem that it is a Wiener process.
That is, we find that $d\overline{Y}_t = dW_t + 2\sqrt{M}\pi_t(F_z)dt$, 
with $W_t$ a Wiener process. 

Furthermore, note that if we start from Eq.\ 
\eqref{eq homodyne Zakai},  
taking $\alpha$ to infinity while 
$M=\alpha^2\kappa^2$ is held constant, 
then we also obtain the 
linear filter Eq.\ \eqref{eq limit linear filter}. 
Likewise, the homodyne filter 
Eq.\ \eqref{eq homodyne filter} 
converges to the filter in Eq.\ 
\eqref{eq limit filter} when 
$\alpha$ is taken to infinity 
while $M=\alpha^2\kappa^2$ is 
held constant.

\section{Decoupling the x-channel}\label{sec elim}

Let us give a brief formal discussion  
to show what happens in the strong driving, 
weak coupling limit. As $\alpha$ increases 
and $\kappa = \sqrt{M}/\alpha$ decreases, 
the relative effect of the atoms on 
the $x$-polarized channel also decreases. 
Therefore, we can reasonably expect that the 
$x$-channel remains in a coherent state. 
Instead of working with respect to the state
$\mathbb{P} = \rho \otimes \phi$ we 
will now work with respect to the state
  \begin{equation*}
  \mathbb{Q} = \rho\otimes \Big\langle \psi^x(f),\ \cdot \ \psi^x(f)\Big\rangle, 
  \end{equation*}
with $f(t) = \alpha e^{i\phi_t}$.    
Note that this means that the $y$-channel is still in 
the vacuum state. Working with respect to 
the coherent state on the $x$-channel means that 
the time evolution is given by Eq.\ \eqref{eq Ut2}.
Formally we can write for $\beta,\gamma \in \{x,y\}$
  \begin{equation*}
  d\Lambda^{\beta\gamma}_t = a^{\beta*}_ta^\gamma_tdt,\ \ \ \ 
  dA^\beta_t = a^\beta_tdt,\ \ \ \ dA^{\beta*}_t = a^{\beta*}_tdt,
  \end{equation*}  
and since for large $\alpha$ and small $\kappa$ 
the $x$-channel is approximately in the coherent 
state $\psi(f)$, we can replace $a^x_t$ by 
$\alpha e^{i\phi_t}$ and $a^{x*}_t$ by $\alpha e^{-i\phi_t}$. 
This means that we obtain for large $\alpha$ and 
small $\kappa$
  \begin{equation*}\begin{split}
  dU^0_t = \Big\{&\big(\cos(\kappa F_z) -1\big) \big(\alpha^2dt + d\Lambda_t^{yy}\big)\ -\\
  &\sin(\kappa F_z)\alpha\big(e^{-i\phi_t}dA_t^y - e^{i\phi_t}dA^{y*}_t\big)\Big\}U^0_t.
  \end{split}\end{equation*}
Now, if we replace $\kappa$ by $\sqrt{M}/\alpha$ and 
take the limit $\alpha$ to infinity, then the 
time evolution satisfies the following QSDE
  \begin{equation*}
  d\overline{U}_t = \left\{\sqrt{M}F_z
  \big(e^{-i\phi_t}dA_t^y - e^{i\phi_t}dA^{y*}_t\big) - 
  \frac{M}{2}F_z^2dt\right\}\overline{U}_t.
  \end{equation*}   
That is, the $x$-channel has been decoupled
from the interaction, see also \cite{JSSP03,GeB06}. 

In a similar way we 
easily see that in the strong driving, weak 
coupling limit we have $dY^+_t = dt$ and 
for $\overline{Y}_t$, the limit of $Y^-_t$, 
we obtain
 \begin{equation}\label{eq limit observations}\begin{split}
 \overline{Y}_t ={ } &\frac{\overline{U}^*_t
 (\Lambda^{xy}_t+\Lambda^{yx}_t)\overline{U}_t}{\alpha} \\
 ={ }  
 &\frac{\overline{U}^*_t
 (\alpha e^{-i\phi_t}A^y_t + \alpha e^{i\phi_t}A^{y*}_t)\overline{U}_t}{\alpha} \\
 ={ }
 &\overline{U}^*_t
 (e^{-i\phi_t}A^y_t + e^{i\phi_t}A^{y*}_t)\overline{U}_t. 
 \end{split}\end{equation}
This shows once more the equivalence of the 
balanced polarimetry experiment and the 
$y$-channel homodyne detection experiment.
It is easy to see that the characteristic 
functional of the process $\overline{Y}$
satisfies the set of coupled differential 
equations of Eq.\ \eqref{eq difeqs}.
Moreover, after decoupling the $x$-channel 
the system is given by $j_t(X) = 
\overline{U}_t^*X\overline{U}_t$ and the 
observations by Eq.\ \eqref{eq limit observations}.
Following essentially the same steps as in 
Section \ref{sec homodyne detection}, we again obtain 
Eqs.\ \eqref{eq limit linear filter} and \eqref{eq limit 
filter} as the linear and normalized 
filters, respectively.

\section{Discussion}

We have provided a quantum stochastic model
Eq.\ \eqref{eq HuP} to describe recent polarimetry 
experiments in which polarized laser light interacts with 
an atomic gas via the Faraday interaction. In 
our description the gauge process plays a 
prominent role. It represents the scattering 
between different channels in the field and it 
provides us with counting processes that can be 
observed. As in \cite{BaL00}, our quantum stochastic 
model presents a novel application to quantum 
optics of the gauge terms in a QSDE. 

Once we set the model, 
we derived quantum filtering equations for 
balanced polarimetry and homodyne detection 
experiments, studied the statistics of 
output processes and obtained filters in the 
strong driving/weak coupling limit.
Our results in the limit confirm the ad hoc filter 
for the balanced polarimetry experiment 
that has already been in use in the 
literature \cite{Sto06}. Moreover, we showed that 
from the point of view of statistical 
inference the balanced polarimetry 
experiment and the homodyne detection 
experiment are equivalent. 

Using formal arguments 
we have seen that in the strong driving, 
weak coupling limit the $x$-channel decouples 
from the description. Rigorous results on this 
decoupling are still to be obtained.

Having an underlying model from which  
rigorous derivations can depart, is likely to be 
advantageous in future further investigations.
In particular, combining the model presented 
in this paper with the results in \cite{YaK03} 
could prove useful for investigating the 
situation where the laser beam passes through 
the gas multiple times \cite{ShM06,SSFMP06}.

\begin{acknowledgments}
L.B.\ thanks Mike Armen, Ramon van Handel and 
Tony Miller for stimulating discussion. L.B. is  
supported by the ARO under Grant W911NF-06-1-0378. 
H.M. and J.S. are supported by the ONR under 
Grant N00014-05-1-0420. H.M. and G.S. are supported 
by the NSF under Grant CCF-0323542.
\end{acknowledgments}

\bibliography{pol}

\end{document}